\documentclass[aps,prb,twocolumn]{revtex4}
\usepackage{graphicx,amssymb}
\begin{document}

\title{Local topological phase transitions in periodic condensed matter systems}
\author{Jan Carl Budich$^1$, and Bj\"orn Trauzettel$^1$} 
\affiliation{$^1$ Institute for Theoretical Physics and Astrophysics,
University of W$\ddot{u}$rzburg, 97074 W$\ddot{u}$rzburg, Germany;}

\date\today
\begin{abstract}
Topological properties of a periodic condensed matter system are global features of its Brillouin zone (BZ). In contrast, the validity of effective low energy theories is usually limited to the vicinity of a high symmetry point in the BZ. We derive a general criterion under which the control parameter of a topological phase transition localizes the topological defect in an arbitrarily small neighbourhood of a single point in $k$-space upon approaching its critical value. Such a local phase transition is associated with a Dirac-like gap closing point, whereas a flat band transition is not localized in $k$-space. This mechanism and its limitations are illustrated with the help of experimentally relevant examples such as HgTe/CdTe quantum wells and bilayer graphene nanostructures.  
\end{abstract}
\maketitle

\section{Introduction}
Over the last three decades topological classification of physical systems has become an important as well as effective tool in condensed matter physics \cite{Volovik,KaneHasan,XLReview2010}. The high interest in the search for experimentally accessible observables represented by topological invariants is due to at least two reasons: First, topology yields a new paradigm of emergent behaviour globally characterizing a bulk system as opposed to microscopic theories describing the dynamics of a system locally by its equations of motion. The rich phenomenology brought about by these new concepts justifies the intense fundamental interest in the field. Second, the possible generic advantages of topologically invariant properties as to future applications say in the field of nanoelectronics are promising \cite{MooreReview}. For the reliability of mesoscopic devices the robustness of properties like a controllably adjustable quantized conductance could be an important prerequisite. An inherent feature of a discrete topological invariant is precisely this robustness against local perturbations of the system as far as they do not break the crucial symmetries entering the global form of the microscopic theory.

We now turn our attention to the topological classification of the Bloch Hamiltonian $H(k)$~the domain of which is the Brillouin zone (BZ) of a periodic solid state system. Due to the periodicity of the bulk system the BZ has the topological structure of a torus. The most typical feature of a topological invariant is that it describes a global feature of the space at hand which is determined by the integral of some characteristic function over the entire BZ. Unfortunately, a microscopic theory describing the solid state system all over the BZ is often times lacking. In such a case, an approved approach to determine e.g. the low temperature/bias transport properties is the construction of a low energy theory with validity regime close to the Fermi surface. For a noninteracting system we think of a $k\cdot p$~theory approximating $H(k)$~as a power series \cite{DresselhausBook}. Without further information about the system, the local approximation for $H(k)$~will in general fail to predict the correct value for the topological invariant when integrated over the BZ. We will substantiate this critical argument using the Integer Quantum Hall Effect (IQHE) \cite{Klitzing1980,Laughlin1981,TKNN1982}
as an example where topology renders all occupied states important as far away from the Fermi surface as they might be. The fingerprints of these states usually considered as inert can then be seen in the low energy transport properties of a finite size system. However, when comparing the system of interest to a well known reference state, the relevant topological features can then be captured by the local approximation for $H(k)$ \cite{BHZ2006}. Generalizing the analysis for a $4+1$-dimensional Dirac Hamiltonian in  Ref. \cite{QiTFT}, we define a rigorous condition under which topological predictions of theories local in $k$-space are conclusive. In this context, we argue that the control parameter $\lambda$~of a topological phase transition can localize the nontrivial features of the transition in $k$-space to any desired accuracy as the control parameter approaches its critical value $\lambda_c$. For a topological insulator, this critical point in $k$-space is the Dirac point where the bulk insulating gap closes. In this case, the global properties of a function on the BZ can be localized in $k$-space without changing the value of the topological invariant. This statement seems to obscure the previous assertion that physically relevant topological properties are essentially global features of the respective system. However, the localization of the topological defect comes at a price: First, the physical robustness of the topological property breaks down, since the protecting mechanism, e.g. the energy gap of a topological insulator goes to zero as $\lambda\rightarrow \lambda_c$. Second, the topological features for the reference state need to be known globally in general to make stringent predictions.

Furthermore, we argue that a phase transition which involves a critical point with flat bands does not satisfy our criterion of a local transition. Finally, we discuss some caveats of the local approach, discussing bilayer graphene as an example where it is well known that symmetries lead to a cancellation of topological defects. In this context, we show that even the topological defect captured by a continuum model of a single valley does not survive a lattice regularization. More subtle effects due to the fermion doubling theorem might lurk if most of the details of a band structure are unknown in more complex materials. Our distinction between a local transition and a nonlocal one is discussed for several prominent examples of topological band structures.

This work is organized as follows: In Section \ref{sec:theory}, we define what we mean by a local topological transition for a generic periodic system. In Section \ref{sec:applications}, we discuss examples which fit into the framework of our definitions and demonstrate that not every transition can be considered local in $k$-space. In Section \ref{sec:summary}, we summarize our results.

\section{Local topological phase transitions}
\label{sec:theory}
Now, we address the mechanism of the localization of a topological defect in $k$-space more precisely for a system of noninteracting electrons in a periodic potential. To do so, we first outline the general structure of a topological transition.
\begin{figure}
\resizebox{0.75\columnwidth}{!}{%
  \includegraphics{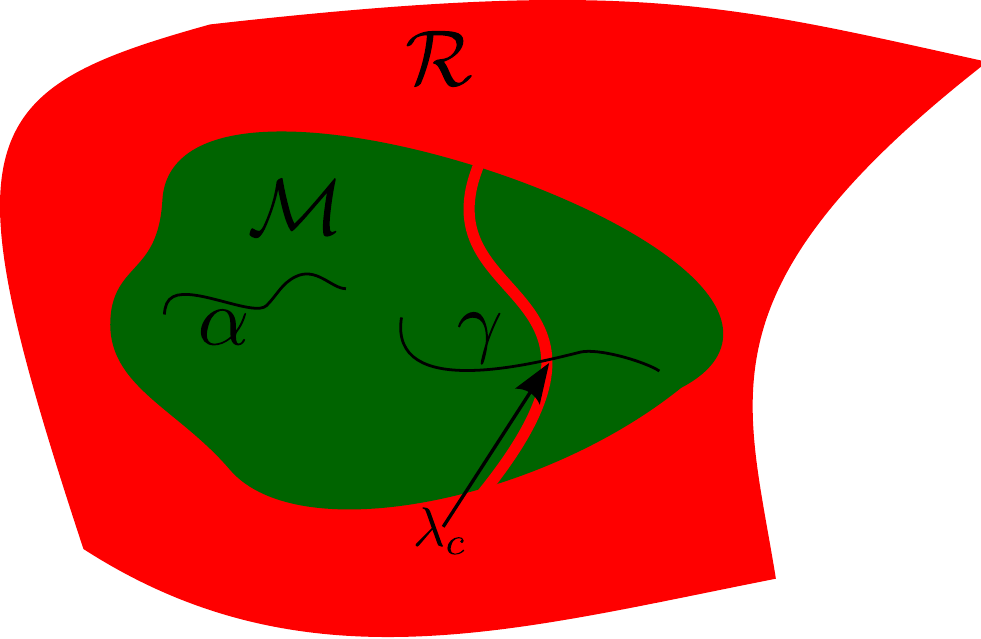}
}
\caption{(Colour online). Parameter manifold $\mathcal R$~(red, light grey) with its submanifold $\mathcal M$~(green, dark grey) on which the homotopy invariant $\mathcal C$~is defined. $\alpha$~is parametrized by the parameter of a homotopy, $\gamma$~by the control parameter $\lambda$~ of a phase transition and crosses between connected components of $\mathcal M$~at the critical point $\lambda_c$.}
\label{fig:parameterspace}
\end{figure}
We assume that the dynamics of the system is governed by an $n$-band Bloch Hamiltonian $H(k,R),~R\in \mathcal R$, where $\mathcal R$~is some parameter manifold parametrizing the set of hermitian complex $n\times n$~matrices respecting the symmetries of the physical system. For the topological invariant $\mathcal C$~to be well defined further conditions, e.g. a bulk insulating gap if the invariant is defined for an insulator, might be necessary. Within these additional constraints the Bloch Hamiltonian is parametrized on the submanifold $\mathcal M\subset \mathcal R$ (see Fig.\ref{fig:parameterspace}). $\mathcal C$~is typically defined as the integral of some characteristic class \cite{MilnorStasheff} over the BZ of the system or over some effective BZ which respects additional symmetries of the system like time reversal invariance (TRI) \cite{Moore2007,FuInversion2007}. As a prominent example the reader might think of the integral of Berry's curvature \cite{Simon1983} over the BZ yielding the first Chern number which represents for example the quantized Hall conductivity $\sigma_{xy}$~in the IQHE regime \cite{TKNN1982,Kohmoto1985} or the contribution of one Kramers partner to the Quantum Spin Hall Effect (QSHE) in HgTe quantum wells as predicted within the Bernevig-Hughes-Zhang (BHZ) model \cite{BHZ2006}. We assume that the Hamiltonian $H(k,\alpha(\lambda))$~smoothly depends on an experimentally controllable parameter $\lambda$~parametrizing a curve $\lambda\mapsto \alpha(\lambda)\in \mathcal M$~(see Fig.\ref{fig:parameterspace}). By construction $\mathcal C$~does not change during the continuous deformation between $H(k,\alpha(0))$~and $H(k,\alpha(1))$.

Let us now think of a topological phase transition in which $\lambda$~is the control parameter upon continuous variation of which the phase transition, i.e. the change of $\mathcal C~$ happens say at $\lambda_c=0$. At $\lambda_c$, the system necessarily leaves the submanifold $\mathcal M$~on which $\mathcal C$~is well defined (see Fig. \ref{fig:parameterspace}). This singularity can be thought of as the closing of an energy gap, for example. Let us assume for simplicity that $\lambda_c=0$~is the only such singularity.

We now analyze the following generic situation ocurring whenever a theory that is local in $k$-space is used to describe topological properties: Assume that we know the function $\mathcal F_{-\delta}(k)$~to be integrated over the BZ for some negative value $\lambda =-\delta$, that is for one value $\mathcal C^-$ of $\mathcal C$. We might think of this regime as some topologically trivial regime, where the features of the system are well known. Predicting the value of the topological invariant $\mathcal C^+$~for positive $\lambda$~amounts to evaluating $\mathcal C^+ =\int_{\mathrm{BZ}}\mathcal F_{+\delta}$, where $\mathcal F_{+\delta}$~is defined for $\lambda =+\delta$.

We are now in a position to formulate a general condition under which a topological phase transition can be successfully described by a low energy theory which is valid in some arbitrarily small neighbourhood $B_\epsilon(K_0)$~of a point $K_0 \in \mathrm{BZ}$.

\begin{figure}
\resizebox{0.95\columnwidth}{!}{%
  \includegraphics[angle=270]{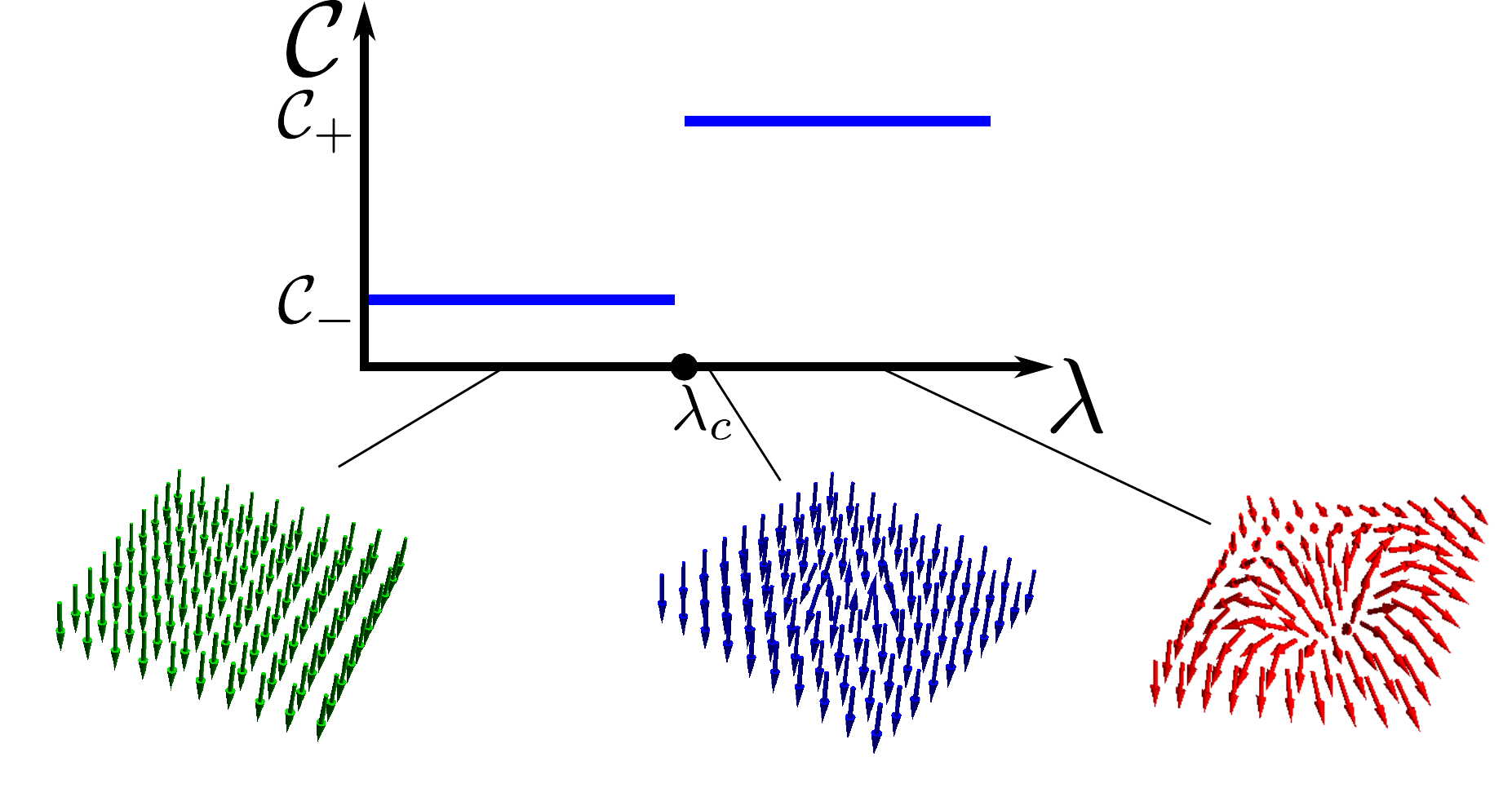}
}
\caption{(Colour online). Sketch of a local transition. The defect is visualized as a pseudo spin configuration for different values of the control parameter $\lambda$. Close to the transition point $\lambda_c$~the difference between the trivial and the nontrivial configuration is localized.}
\label{fig:localTrans}
\end{figure}

{\bf{Local transitions}}\\
{\it{A topological phase transition is called local in $k$-space at $K_0$~if}}
\begin{eqnarray}
\lim_{\delta\rightarrow0^+}\int_{\mathrm{BZ}\setminus B_\epsilon \left(K_0\right)}\left(\mathcal F_{+\delta}-\mathcal F_{-\delta}\right)=0.
\label{eqn:localTrans}
\end{eqnarray}
for any validity range $\epsilon$~of the low energy model.
The meaning of this criterion is readily illustrated (see also Fig. \ref{fig:localTrans}).
During the phase transition the discrete value of $\mathcal C$~has changed by some finite amount, say $N$. This implies
\begin{eqnarray}
\int_{\mathrm{BZ}}\left(\mathcal F_{+\delta}-\mathcal F_{-\delta}\right)=N,~\delta>0
\label{eqn:rapidTrans}
\end{eqnarray}
The combination of Eq. (\ref{eqn:localTrans}) and Eq. (\ref{eqn:rapidTrans}) yields the following expression for the desired quantity $\mathcal C^+$:
\begin{eqnarray}
&\mathcal C^+=\int_{\mathrm{BZ}}\mathcal F_{+\delta}=\lim_{\delta\rightarrow0^+}\int_{\mathrm{BZ}}\mathcal F_{+\delta}=\nonumber\\%\overset{(\ref{eqn:localTrans})}{=}\nonumber\\
%&\lim_{\delta\rightarrow0^+}\left(\int_{\mathrm{BZ}}\mathcal F_{-\delta}+\int_{B_\epsilon \left(K_0\right)}\left(\mathcal F_{+\delta}-\mathcal F_{-\delta}\right)\right)=\nonumber\\
&C^-+\lim_{\delta\rightarrow0^+}\int_{B_\epsilon(K_0)}\left(\mathcal F_{+\delta}-\mathcal F_{-\delta}\right)
%\overset{(\ref{eqn:rapidTrans})}{=}
=N+\mathcal C^-
\end{eqnarray}
Note that a phase transition classified as local in $k$-space requires the knowledge of $\mathcal F$~only in an arbitrarily small neighbourhood of a point $K_0$~to calculate $\mathcal C$~on the topologically nontrivial/unknown side of the transition point. Therefore, in a situation satisfying the above criterion, the change of the topological defect can be localized within the validity regime of the given model by tuning a knob experimentally. Only if this is the case the global topological properties of the low energy model have stringent physical implications given the fact that it describes the physical system only locally around $K_0$. The above criterion can easily be generalized to include a transition which happens locally around several points $K_1,\ldots, K_n~$ in the BZ. In this case, a low energy theory valid around each of these points has to be known to make conclusive predictions about the change of $\mathcal C$.

\section{Examples and counter-examples}
\label{sec:applications}
We now turn to examining several experimentally relevant examples of solid state systems with interesting $k$-space topology. Our examples will show how efficiently our general analysis can be applied to classify topological phase transitions as to their locality and to detect subtle problems that might preclude assigning rigorously defined topological invariants to $k.p$-models not satifying our criterion.

\subsection*{QSHE in HgTe/CdTe quantum wells}
In a seminal paper \cite{Avron1988}, Avron {\it{et al.}} showed that topological defects in TRI halfinteger fermion systems may well exist. These authors point out that a single Kramers partner can have a nonvanishing Chern number as observed in the IQHE regime whereas the Kramers pair's total Chern number obeys a zero sum rule. A realization of this phenomenology is the QSHE \cite{KaneMele2005a,KaneMele2005b,BHZ2006}, where the relevant $\mathbb Z_2$~topological invariant $\nu$ can be thought of as the difference between these two contributions with opposite sign modulo two. Let us consider the BHZ model for the QSHE \cite{BHZ2006}.
The BHZ Hamiltonian defining a low energy theory around the $\Gamma$-point, i.e. around $k=0$, reads
\begin{eqnarray}
H(k)=\left(\begin{array}{cc} {h(k)}&0\\0&{h^*(-k)} \end{array}\right)
\label{eqn:bhzham}
\end{eqnarray} 
with $h(k)=\epsilon(k)\sigma_0+d_i(k)\sigma_i$,
$d_1+id_2=A(k_x+ik_y),~d_3=M-B(k_x^2+k_y^2)$, 
where the gapless part $\epsilon(k)$~turns out to be irrelevant for the topological classification. This model is a $k\cdot p$-type perturbative expansion around the $\Gamma$-point up to second order in $k$. Let  us assume that the validity regime of this theory is at least $B_\epsilon(\Gamma)$. The Kramers Chern number defined for each of the blocks is given by \cite{Qi2006}
\begin{eqnarray}
N_K=\int_{\mathbb R^2}d^2 k \mathcal F(k)
\label{eqn:KramersChern}
\end{eqnarray}
where $\mathcal F(k)\equiv \frac{1}{4 \pi}\hat d(k)\left(\partial_{k_x} \hat d(k)\times \partial_{k_y}\hat d(k)\right)$. The unit vector $\hat d = \frac{\vec d}{| \vec d|}$~is well defined for all values of $k\in \mathbb R^2$~if $M\ne 0$, otherwise it has a single singularity at the $\Gamma$-point. In order for $N_K$ to be well defined as a Chern number, there are two formally possible ways to compactify the $k$-space $\mathbb R^2$~of the 2D continuum model to a non contractible manifold \cite{BHZ2006}. As $\hat d(k\rightarrow \infty)=-\mathrm{sgn}(B)\hat e_z$~does not depend on the polar angle of $k$~one can compactify $\mathbb R^2$~to a sphere $S_2$~by identifying $\infty$~with the northpole \cite{Volovik}, without loosing information about the map $\hat d$. Another way to go would be a lattice regularization of the continuum model which provides the $k$-space with the topology of a torus $T^2$, i.e. introducing a BZ. This does not change the effective Hamiltonian in its validity regime. The topological phase transition happens at $M=0$, the control parameter tuning $M$~is the thickness of the quantum well $t$~with critical value $t_c$. Thus $t$~is what we called the control parameter $\lambda$~in our general analysis in Section \ref{sec:theory}. We now analyze  this transition regarding our criterion (\ref{eqn:localTrans}). For $M=\pm \tilde \delta$~small that is $t=t_c\pm \delta$, the terms linear in k will dominate the constant contribution $M$ for any finite value of k. Thus as $\delta\rightarrow 0^+$~we calculate
\begin{eqnarray}
1=\int_{T^2}\left(\mathcal F_{+\delta}-\mathcal F_{-\delta}\right)=\int_{B_\epsilon(\Gamma)}\left(\mathcal F_{0^+}-\mathcal F_{0^-}\right).
\end{eqnarray}
In this sense, the transition predicted by the BHZ model is local at the $\Gamma$-point in the language of our criterion (\ref{eqn:localTrans}).

Beyond the physics described by the 4-band model in Eq. \ref{eqn:bhzham}, a critical discussion of the theoretical prediction of the QSHE thus reduces to examining two points. First, the possibility should be ruled out that due to unidentified symmetries some other critical point in the Brillouin zone emerges at the same thickness $t_c$, e.g. due to another level crossing~which would then not be comprised in the theory and could cancel the defect at $k=0$. Second, since the predicted QSHE becomes experimentally relevant at finite energy gap $M_e$, that is at $t_e>t_c$~the existence of further critical points at $t_c<t<t_e$~should be forbidden. Both criteria are fulfilled in HgTe/CdTe quantum wells \cite{Konig2007}.
\subsection*{A first counter-example regarding HgTe/CdTe quantum wells}
We now show that for the BHZ model criterion (\ref{eqn:localTrans}) is not met independently of the physical situation. To understand this let us assume that the experimentalist could tune the parameter $B$~instead of the now fixed parameter $M=1$, i.e. $B$~is tuned by the control parameter $\lambda$~occurring in our general analysis in Section \ref{sec:theory}. The continuum model would predict
\begin{eqnarray}
\int_{\mathbb R^2}\mathcal F=\frac{\mathrm{sgn}(B)+1}{2}
\end{eqnarray}
However, the transition at $B=0$~happens at $k\rightarrow \infty$~far away from the $\Gamma$-point and can thus not be considered as a local transition. If we now compactify the base space by a lattice regularization with lattice constant $a=1$~the prediction for the topological invariant is
\begin{eqnarray}
\int_{T^2}\mathcal F=\theta\left(B-\frac{1}{8}\right),
\end{eqnarray}
where $\theta$~denotes the heavyside step function. This describes a phase transtition at $B=\frac{1}{8}$~which is local at $k=(\pi,\pi)$, again far away from the $\Gamma$-point, where the model is valid. We conclude that given this modified physical situation the predictions of the model should be taken with great care, since the lattice regularization shifts the critical point of the phase transition as well as its location in k-space. This in turn implies that the critical point can depend on the way the lattice is regularized.
\subsection*{IQHE with degenerate Landau levels as a nonlocal defect}
In both examples mentioned up to now, analytical progress in the topological classification could be made with the help of a low energy theory. In the following, we argue why the situation is in principle different for a system with an integer number of filled Landau levels, i.e. a system in the IQHE regime. The failure of a low energy theory is in this case already indicated by the following observation. Consider an IQHE system at low temperature $T$~and low bias voltage $V$. One might expect that a theory describing the system in a sufficiently large neighbourhood $\Delta E$~of its Fermi energy as measured by the energy scale $\mathrm{max}(T,V)$~should be capable of predicting low energy transport properties of the sample. However, in this case, topology renders all occupied states important, since the edge states consist of equal contributions from all occupied bands. It has been shown \cite{Kohmoto1985} that the topological defects in the BZ leading to the nonvanishing Chern number representing $\sigma_{xy}$~are centered around the zeros of the Bloch function $k\mapsto \langle r_0| u_k\rangle=u_k(r_0)~$ for fixed $r_0$. However, the position of these zeros in the BZ depends on the choice of $r_0$~which in turn amounts to fixing a gauge on the corresponding $U(1)$-fiber bundle over the BZ with fiber $\left[| u_k\rangle\right]\equiv\left\{e^{i\phi}| u_k\rangle| e^{i\phi}\in U(1)\right\}~$ at $k$. Therefore, the position of these defects in $k$-space will change as the gauge is switched. Given the fact that physical observables are gauge invariant, this localization cannot be performed by some experimentally accessible control parameter. This example fits into our line of reasoning demonstrating that a transition between flat bands, here represented by the filled Landau levels is nonlocal in $k$-space as opposed to a transition associated with a critical Dirac gap closing point. 
\subsection*{A defect that cannot be compactified regarding single valley bilayer graphene}
Next, we study an example of a globally topologically trivial system, namely gapped bilayer graphene, where a cancellation of two topological defects each of which is local at an individual high symmetry point in the BZ takes place. Beyond this well known phenomenology, we will show that even the effective low energy Hamiltonian governing the system around one of the valley points $K,K'$, gives rise to a topologically trivial 2D band structure in the sense of a rigorous definition. Labelling the individual valleys by $\tau=\pm~$ the effective Hamiltonian can be written as \cite{LiPRB2010}
\begin{eqnarray}
H_\tau(k)=-d_\tau^i \sigma_i
\label{eqn:BLGHam}
\end{eqnarray}
with $d^1_\tau=k_x^2-k_y^2,~d_\tau^2=2\tau k_x k_y,~d_\tau^3=\Delta$. In the validity regimes of these low energy Hamiltonians, the function $\mathcal F$, whose integral over the whole BZ yields the Chern number $\mathcal C$~is again given by $\mathcal F^\tau(k)\equiv \frac{1}{4 \pi}\hat d_\tau(k)\left(\partial_{k_x} \hat d_\tau(k)\times \partial_{k_y}\hat d_\tau(k)\right)$. This function does not vanish around either of the high symmetry points so that a local topological phase transition at the gap closing point $\Delta =0$~could be suspected at first glance. However, since $\mathcal F^+\equiv-\mathcal F^-$, we can conclude that the total contribution of both valleys $\tau=\pm$~will always vanish regardless of the sign of $\Delta$, rendering the global topology of the BZ trivial. Nevertheless, something nontrivial happens to exist also in this case. If we artificially confine our attention to a single valley $\tau$~and investigate the behaviour of $\mathcal F$~when crossing the critical point $\Delta =0$, we observe $\mathcal F_\Delta^\tau=-\mathcal F_{-\Delta}^\tau$, which might indicate a local phase transition at a single $K$-point. As $\Delta\rightarrow 0^+$, the support of $\mathcal F^\tau_{\pm \Delta}$~can be localized in the vicinity of $K$~to any accuracy, that is within any $B_{\epsilon}(K)$, albeit 
\begin{eqnarray*}
\forall_{\epsilon>0}\lim_{\Delta\rightarrow 0^+}\int_{B_\epsilon(K)}\left(\mathcal\mathcal F^\tau_\Delta-\mathcal F^\tau_{-\Delta}\right)=2
\end{eqnarray*}
which looks as if the Chern number changes by two in a local transition in the language of criterion (\ref{eqn:localTrans}). However, discussing the BHZ model, we pointed out that the Chern number is only well defined when the $k$-space of the system is compactified. Otherwise algebraic topology tells us that all characteristic classes must be trivial since the $k$-space of the continuum model is contractible \cite{ChoquetBruhat}. For the BHZ model we explicitly showed how the topological defect of $\hat d$~is unaltered by this compactification for the phase transition at $M=0$.
\begin{figure}
\centering
\resizebox{0.65\columnwidth}{!}{%
  \includegraphics{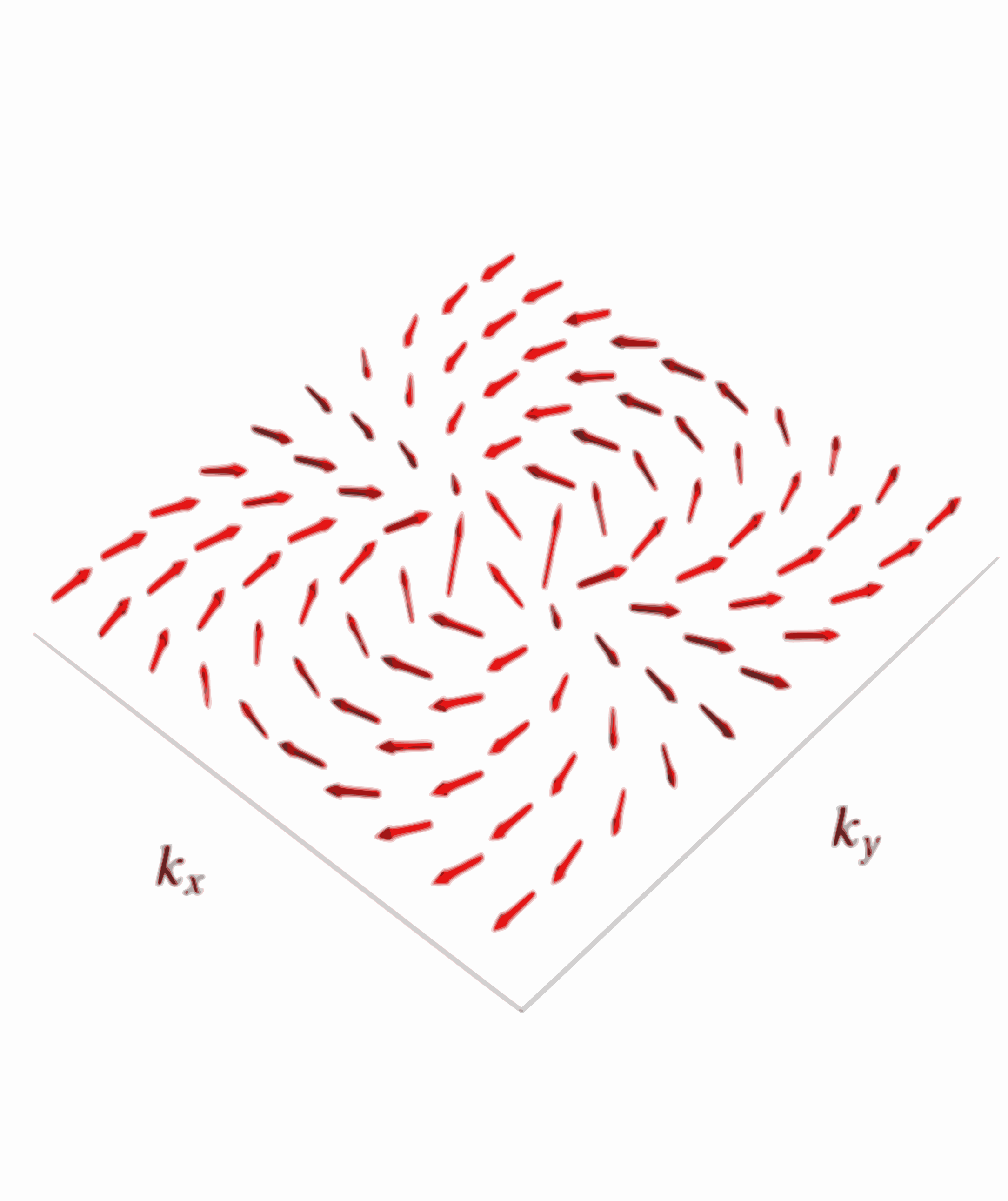}
}
\caption{(Colour online). Vorticity of $\hat d_\tau(k)$~for $\Delta=1$. The configuration is a double covering of the hemisphere so that $k$-space cannot be compactified comprising the topology of this configuration.}
\label{fig:defect}
\end{figure}

The vorticity of the configuration $\hat d_\tau $ (see Fig. \ref{fig:defect}) cannot be encompassed by a compactified $k$-space. This gives a good intuition as to why for any lattice regularization (index $c$), e.g. $d^1_{c,\tau}(k)=2(\cos{k_y}-\cos{k_x}),~d^2_{c,\tau}(k)=2\sin{k_x}\sin{k_y},~d^3_{c,\tau}=\Delta$, where we have chosen unit lattice constant for simplicity, the integral
\begin{eqnarray}
\int_{T^2}\mathcal F_\Delta^{c,\tau} = 0,
\end{eqnarray}
i.e. any properly defined Chern number vanishes. This means that bilayer graphene, as described by the Hamiltonian (\ref{eqn:BLGHam}), has a topologically trivial 2D bandstructure. We have shown that this statement holds even if we just consider the Hamiltonian of one valley and do not take into account the existence of the other valley. The deep physical reason for this behaviour stems from the fermion doubling theorem \cite{Nielsen1981} which already predicts that a second defect has to exist which can add to or as in this case cancel the contribution of an individual valley.

\section{Summary}
\label{sec:summary}
To summarize, we carefully investigated the question to which extent low energy theories that are local in $k$-space can be used to successfully capture the topology of the Bloch Hamiltonian on the BZ. We found a both illustrative and general criterion to classify when a topological phase transition can be considered local in the BZ. With the help of this criterion, we could analyze the rigorous conditions under which existing models can make stringent experimental predictions. Our analysis is valid independently of the dimension of the model system. Many of our statements can be straightforwardly generalized to interacting systems by formulating the topological invariants in terms of the single particle Green functions as far as the interacting system is adiabatically connected to the noninteracting case \cite{Volovik}. We demonstrated the meaning of our predictions for three different examples: the quantum spin Hall effect, the integer quantum Hall effect, and bilayer graphene.
\section*{Acknowledgment}
We thank the DFG-JST Research Unit "Topological Electronics" for financial support and Dietrich Rothe as well as Patrik Recher for fruitful discussions.
% \bibliographystyle{unsrt}
% \bibliography{letopoiop}

\end{document}